\documentclass[aps,prb,twocolumn,showpacs,amssymb,superscriptaddress]{revtex4-1}

\usepackage{amsmath}  
\usepackage{amsfonts} 
\usepackage{graphicx} 

\usepackage{color}			

\usepackage[pdftex,colorlinks=true,pdfstartview=Fit]{hyperref}		

\begin{document}


\title{From rods to blobs:  When geometry is irrelevant for heat diffusion}



\author{Matthew Frick}
\email{mfrick@sfu.ca}
\affiliation{Department of Physics, Simon Fraser University, Burnaby, B.C., V5A 1S6, Canada}

\author{Swapnil Gupta}
\email{swapnilgupta.229@gmail.com}
\affiliation{Department of Physics, Simon Fraser University, Burnaby, B.C., V5A 1S6, Canada}

\author{John Bechhoefer}
\email{johnb@sfu.ca}
\affiliation{Department of Physics, Simon Fraser University, Burnaby, B.C., V5A 1S6, Canada}

\date{\today}

\begin{abstract}
Thermal systems are an attractive setting for exploring the connections between the
lumped-element approximations of elementary circuit theory and the partial-differential field equations of mathematical physics, a topic that has been neglected in physics curricula.  In a calculation suitable for an undergraduate course in mathematical physics, we show that the response function between an oscillating heater and temperature probe has a smooth crossover between a low-frequency, ``lumped-element" regime where the system behaves as an electrical capacitor and a high-frequency regime dominated by the spatial dependence of the temperature field.  Undergraduates can easily (and cheaply) explore these ideas experimentally in a typical advanced  laboratory course.  Because the characteristic frequencies are low, ($\approx$ 30~s)$^{-1}$, measuring the response frequency by frequency is slow and challenging; to speed up the measurements, we introduce a useful, if underappreciated experimental technique based on a multisine power signal that sums carefully chosen harmonic components with random phases.  Strikingly, we find that the simple model of a one-dimensional, finite rod predicts a temperature response in the frequency domain that closely approximates experimental measurements from an irregular, blob-shaped object.  The unexpected conclusion is that the frequency response of this irregular thermal system is nearly independent of its geometry, an example of---and justification for---the ``spherical cow" approximations so beloved of physicists.

\end{abstract}

\maketitle

\section{Introduction} 

In their first year of undergraduate physics, students generally work with ``lumped" objects:  rigid bodies in mechanics; uniformly cooling objects in thermodynamics (Newton's law of cooling); resistor-capacitor-inductor circuits in electricity and magnetism.  In intermediate courses, these topics are revisited:  rigid bodies become elastic, temperatures diffuse, and charged objects create interesting electric fields.  Traditionally, little effort has been made to show how lumped-element approximations arise from the underlying field equations.  In the engineering and applied-physics literature, discussions are more common.\cite{bergman11,ott88}  Perhaps this disconnect occurs because in two of the common settings, mechanics and electromagnetism, the calculations that connect field equations to lumped elements in a quasistatic limit can be relatively complex.\cite{ramo94,smith97}  Indeed, even graduate-level texts that discuss the quasistatic limit often treat the reduction from Maxwell's equations to AC circuits only qualitatively\cite{zangwill13} or very briefly.\cite{landau84} 

Here, we show that thermal problems, which depend on a scalar temperature field, are a much simpler setting in which to explore such connections.  In addition, the central physical quantity in such problems, the thermal diffusivity $\alpha$, is often interesting on its own, and techniques for measuring diffusivity continue to be developed and play important roles in experimental physics research\cite{mandelis00,ogi16,zhang17} and in teaching physics.\cite{bodas98,sullivan08,anwar14}   A recent article in this journal on the temperature response of a heat fin focuses on the time-domain step response, which is complementary to the frequency-domain approach taken here.\cite{brody17}  Although frequency response is perhaps more abstract, it is simpler to describe mathematically.

We consider heat propagation along a one-dimensional rod, showing that in the low-frequency limit, the problem reduces to one of a thermal ``capacitor" and in the high-frequency limit to thermal waves in a semi-infinite medium.  A perhaps-surprising observation is that the limiting behaviors are largely independent of the geometry of the system.  We confirm this conclusion experimentally on an irregularly shaped object, using techniques for efficiently measuring the linear response of systems with sluggish dynamics.  The results give some insight into the reasons that ``spherical cow" approximations can be surprisingly successful.

\section{Heat diffusion along a rod}
\label{sec:heatDiffusion}

\begin{figure}
	\includegraphics[width=1.8in]{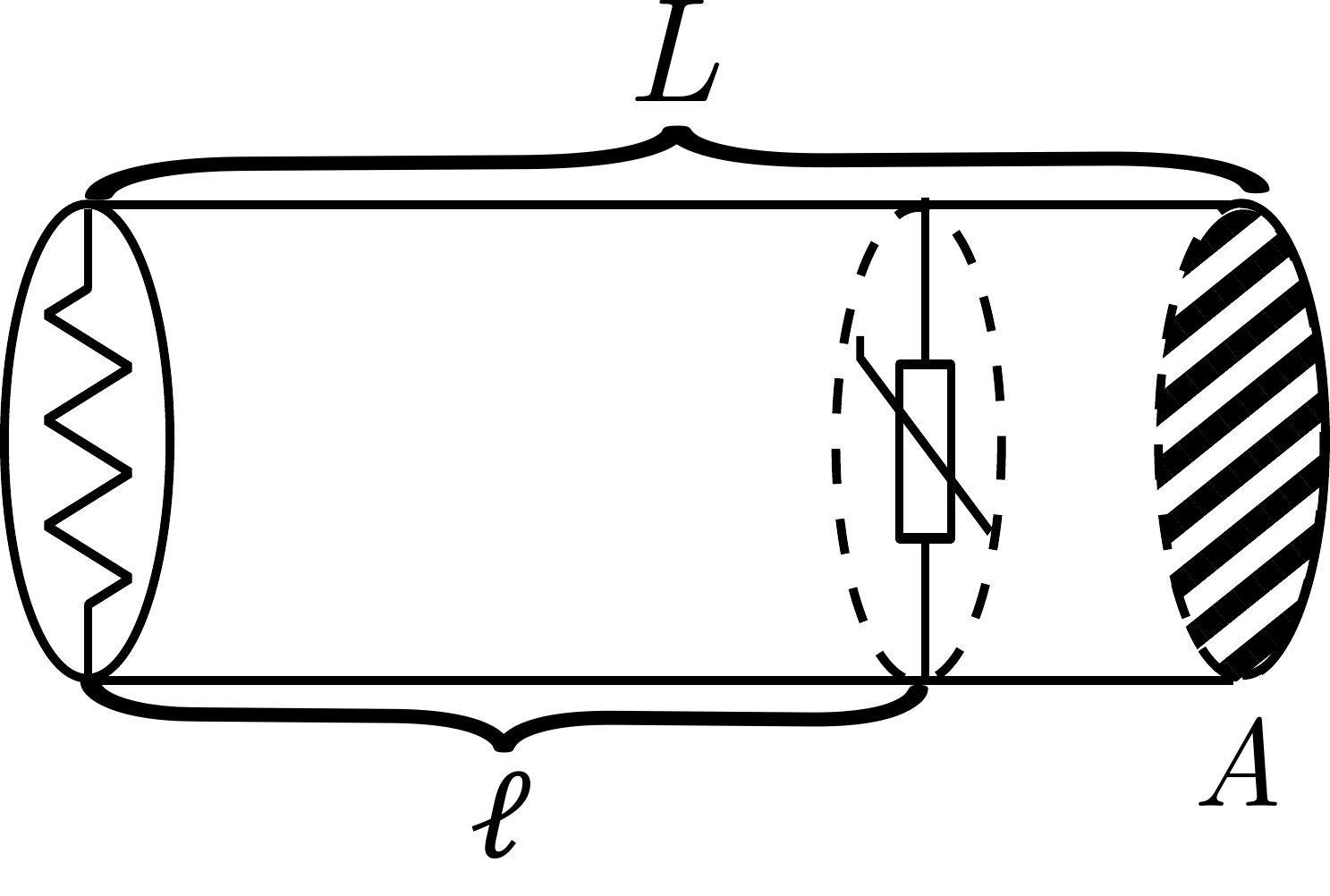}
	\caption{Model for heat flow:  a one-dimensional rod of length $L$ and cross-section $A$ has a heater at $x=0$ and temperature probe at $x=\ell$.}
\label{fig:Model}
\end{figure}

We introduce a one-dimensional model of heat diffusion that has simple, instructive limits for both low and high frequencies.  Consider a long rod of length $L$ and cross-sectional area $A$ (Fig.~\ref{fig:Model}).  At one end ($x=0$) is a heater that injects power.  Along the rod is a temperature probe, at a distance $x=\ell < L$.  The rod has thermal diffusivity $\alpha = \lambda/ (\rho c_p)$, with $\lambda$ the thermal conductivity, $\rho$ the density, and $c_p$ the specific heat capacity at constant pressure.

We model heat flow in the rod by a diffusion equation with distributed losses $\mu$ to the surrounding environment,\cite{brody17}
\begin{equation}
	\frac{\partial T}{\partial t} = \alpha \frac{\partial^2 T}{\partial x^2}-\mu \, T(x,t) \,,
\label{eq:diffusion}
\end{equation}
where the one-dimensional temperature field $T(x,t)$ is measured relative to the ambient  environment.  In our investigation, we will use periodic power signals to measure the parameters $\alpha$ and $\mu$ in Eq.~\eqref{eq:diffusion}, an idea that traces back to early work by $\AA$ngstr\"om (1861), who carried out the first successful measurement of the thermal diffusivity of copper.\cite{angstrom1861} 

For periodic power injection, the boundary conditions at $x=0$ and $L$ are given by 
\begin{equation}
	-\lambda\left.\frac{\partial \tilde{T}}{\partial x}\right|_{x=0} = \frac{\tilde{P}(\omega)}{A}
		\quad \text{and} \quad
	\left.\frac{\partial \tilde{T}}{\partial x}\right|_{x=L} = 0 \,,
\label{eq:BCs}
\end{equation}
where $\tilde{P}(\omega)/A$ is the Fourier transform of the heat flux produced by the heater at $x=0$ and where the rod is insulating at $x=L$.  The temperature field is $\tilde{T}(\omega,x)$.

\begin{figure}
\includegraphics[width=3.0in]{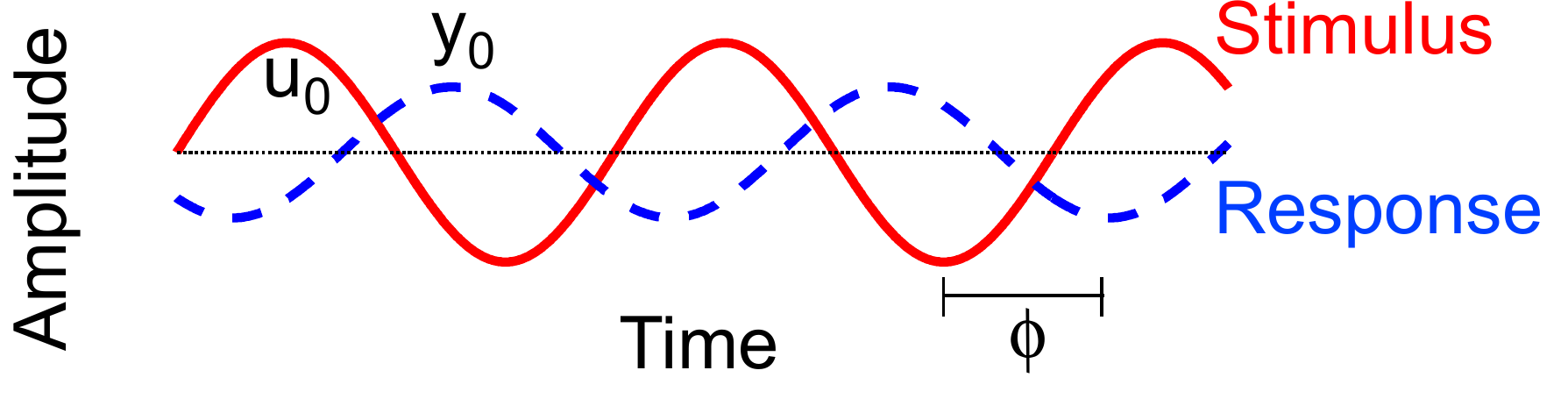}
\caption{Sinusoidal power signal $P(t)$ leads to a sinusoidal temperature response $T(\ell,t)$ at the probe site $x=\ell$.  The magnitude of the dynamical response $\tilde{\chi}$ corresponds to the relative amplitude of the two signals and the argument to the phase shift $\phi$, as shown.}  
\label{fig:SingleSine}
\end{figure}

Our problem is closely related to standard solutions of the diffusion equation.\cite{carslaw59}  In  Appendix~\ref{appendix:responseDerivation}, we Fourier transform Eq.~\eqref{eq:diffusion} in time and solve the second-order differential equation for $x$ to find the linear response between heater and probe,
\begin{equation}
	\tilde{\chi}_\ell(\omega) \equiv \frac{\tilde{T}(\omega)}{\tilde{P}(\omega)} 
		= \left( \frac{1}{A\lambda} \right) \, \frac{\cosh [ a (L-\ell) ]} {a \, \sinh\left( a L \right)} \,,
\label{eq:FullModel}
\end{equation}
where $a = a(\omega) \equiv \sqrt{(\mu-\text{i}\omega)/\alpha}$.  Equation~\eqref{eq:FullModel} gives the relative amplitude and phase of the temperature at $x=\ell$ for an AC power input $\tilde{P}(\omega)$  at angular frequency $\omega$.  See Fig.~\ref{fig:SingleSine}.

To understand better the dynamical response, we examine its low- and high-frequency limits.  Since the dimensions of diffusivity $\alpha$ are $\bigl($length$^2$/time$\bigr)$, we can see that $\ell^2/\alpha$ is the time for a temperature variation to diffuse from the heater to the probe.   In Appendix~\ref{appendix:engVphys}, we contrast this kind of physical approach to understanding response-function dynamics with the more phenomenological approach taken by engineers.

\subsection{Low-frequency limit}
\label{sec:lowfreq}

Let $\omega \ll \alpha/\ell^2$, implying $|a(L-\ell)| \ll 1$ and $|aL| \ll 1$, so that $\cosh (\cdot) \approx 1$ and $\sinh aL \approx aL$.  In this limit, 
\begin{equation}
	\tilde{\chi}_\ell(\omega) \approx \frac{1}{A L \lambda a^2} 
	= \frac{\alpha}{AL\lambda (\mu-\text{i}\omega)} \,,
\label{eq:LFModel}
\end{equation}
which represents a low-pass filter with loss rate $\mu$ to the environment.  Notice that the volume of the rod $V=AL$ is ``assembled" from the cross-sectional rod area $A$ that is a prefactor in Eq.~\eqref{eq:FullModel} and the rod length $L$ that comes from the sinh term.  We use the volume to define the total mass of the rod, $M = V\rho$.  The heat capacity $C \equiv AL \lambda / \alpha =  AL \rho c_p$, is the product of the specific heat and sample mass.  Thus, in this low-frequency limit, the geometry of the sample enters only indirectly. 

If the rod is insulated from the environment, the losses can be small.  In the limit $\mu \ll \omega \ll \alpha/\ell^2$, the response function further simplifies to
\begin{equation}
	\tilde{\chi}_\ell(\omega) \approx \frac{\alpha}{AL\lambda (-\text{i}\omega)} 
		\equiv \frac{1}{-\text{i}\omega C} \,,
\label{eq:lowfreq-Cap}
\end{equation}
which is identical to that of an electrical capacitor.

\subsection{High-frequency limit}
\label{sec:highfreq}

When $\omega \gg \alpha/\ell^2$, we can set $\mu = 0$, since we have assumed that $\alpha/\ell^2 \gg \mu$.  Then, for $\theta \gg 1$, one can approximate $\sinh \theta \approx \cosh \theta \approx \tfrac{1}{2} {\rm e}^{\theta}$, and the dynamical linear-response function becomes
\begin{equation}
	\tilde{\chi}_\ell(\omega) \approx \frac{1}{A \lambda} \, 
		\frac{{\rm e}^{a(L-\ell)}}{a \, {\rm e}^{aL}} 
	=  \left( \frac{1}{A \lambda} \right) \frac{{\rm e}^{-a\ell}}{a}   \,,
\label{eq:HFModel}
\end{equation}
where $a(\omega) \approx \sqrt{-\text{i}\omega/\alpha}$.  Let us define a length $\ell_0 \equiv \sqrt{2\alpha /\omega}$, so that $a = \sqrt{-2\text{i}}/\ell_0 = (1-\text{i})/\ell_0$.  Thus,
\begin{equation}
	\tilde{\chi}_\ell(\omega) \approx 
	\left( \frac{\ell_0}{A \lambda} \right) \,
		\left( \frac{{\rm e}^{\text{i} \ell / \ell_0} }{1-\text{i}} \right) \, {\rm e}^{-\ell / \ell_0} \,.
\label{eq:HFModel1}
\end{equation}
Equation~\eqref{eq:HFModel1} shows that temperature oscillates and decays in space, with a period and decay length of $\ell_0$.  Notice that $\tilde{\chi}_\ell(\omega)$ is now independent of the rod length $L$.  In this limit, the decaying thermal waves are analogous to results for the ``skin depth" of alternating EM fields in conductors.\cite{zangwill13}

\subsection{Geometry and limits}
\label{sec:geometry}

\begin{figure}
	\includegraphics[width=3in]{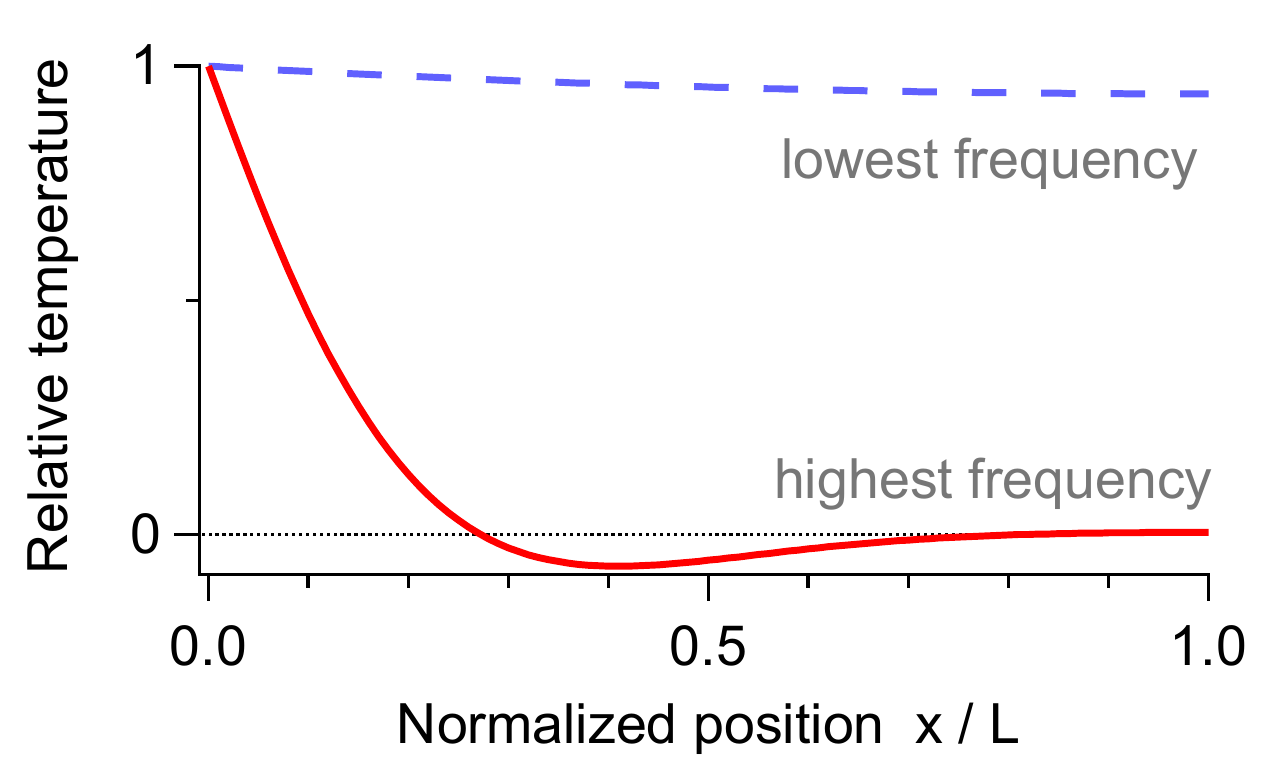}
	\caption{Spatial temperature profile of the model system in the low-frequency (dashed line) and high-frequency (solid line) limits. The low-frequency data is the lowest frequency considered in our fits (1 mHz) and is constant to $\approx 6\%$. The high-frequency data likewise correspond to the highest frequency considered in our fits (2 Hz), and the temperature oscillations are heavily damped, being very close to zero starting at four fifths the rod length.}
\label{SpatialProfile}
\end{figure}

The striking feature of both limits, Eqs.~\eqref{eq:lowfreq-Cap} and \eqref{eq:HFModel1}, is that the geometry of the rod has disappeared from the final expression.  The key physical point is that oscillations decay away from a source over a length scale $\ell_0$.  In the low-frequency limit, $\ell_0 \gg L$, and the temperature is approximately constant over the sample, as illustrated in Fig.~\ref{SpatialProfile}.  An oscillating power input then leads to an approximately uniform temperature oscillation whose amplitude depends on the sample mass.  The geometry enters via the volume but is otherwise unimportant.  Such a limit is sometimes known as the lumped-element approximation.  It applies, as well, to the components of the electrical circuits treated in elementary physics courses.

In the high-frequency limit, $\ell_0 \ll L$, and the temperature fluctuations decay to zero well before reaching the other end of the rod, at $x=L$, as can also be seen in Fig.~\ref{SpatialProfile}.  The rod becomes effectively semi-infinite, and only the distance $\ell$ between heater and probe is important.  This type of geometry is often studied in intermediate-level mathematical physics courses.  

In the simple setting of one-dimensional temperature diffusion, we easily see how both limits arise from the general linear-response function.  Similar limits exist in an electromagnetic setting, starting from Maxwell's equations, but the calculations are notably more involved.

\section{Experimental tests}
\label{sec:experiment}

\begin{figure}[h]
\includegraphics[width=3.4in]{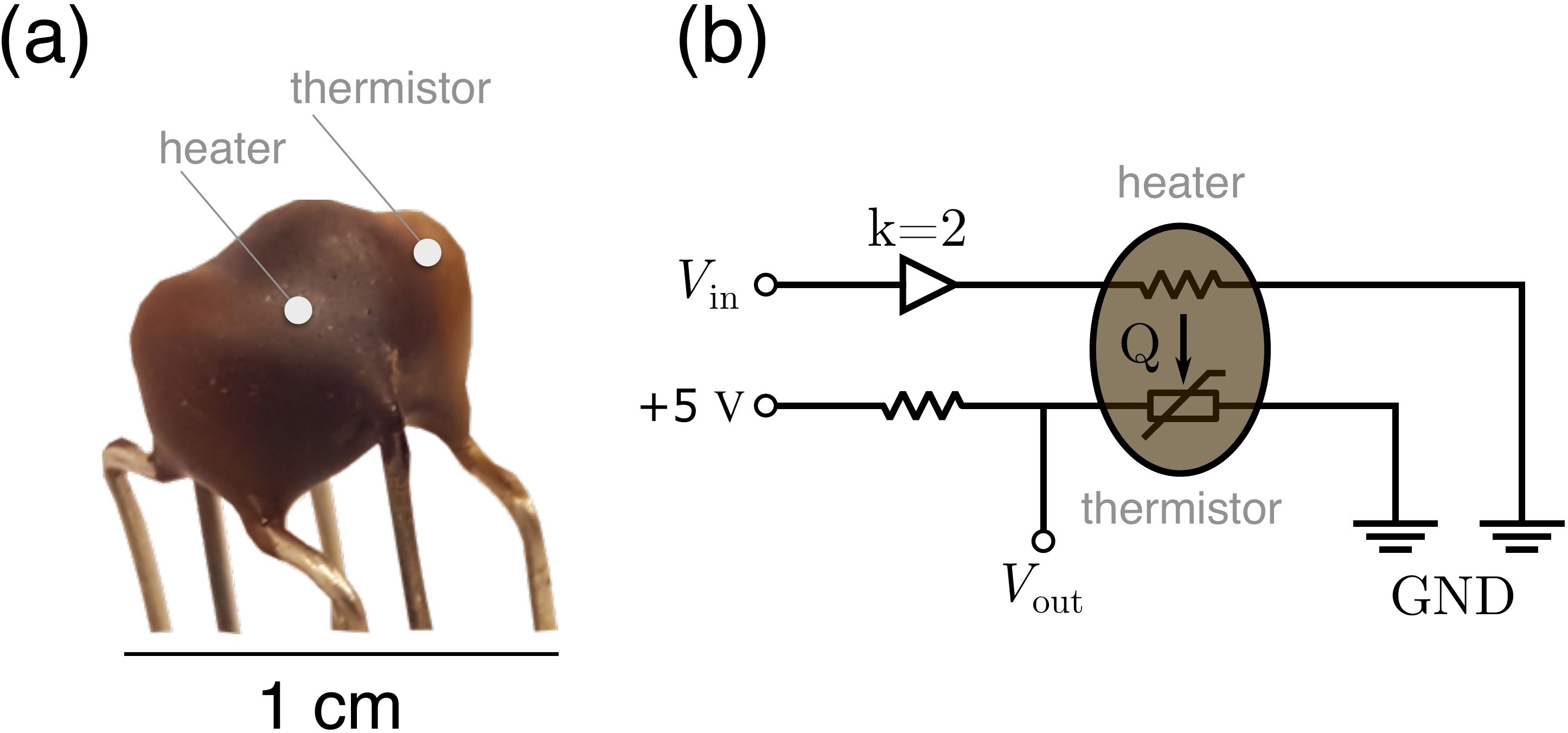}
\caption{Experimental setup. (a) ``Blob shaped'' epoxy encases a heater resistor and two thermistors (only one used).  (b) Schematic diagram, including data-acquisition circuits.  Voltage from the D/A converter, $V_{\rm in}$ is sent to a power amplifier with gain of two.  Heat $Q$ flows through the epoxy to a thermistor in a voltage-divider circuit.}
\label{fig:bug}
\end{figure}

In Section~\ref{sec:heatDiffusion}, we showed that the low- and high-frequency limits of temperature diffusion along a rod led to expressions that were independent of the details of the rod's geometry.  In both cases, the physical reasoning suggests that such independence should hold for a wide variety of geometries.  To test this idea, we investigated experimentally heat propagation in the irregular, ``blob shaped" system depicted in Fig.~\ref{fig:bug}a.  

The experimental system consists of a 27.6-$\Omega$ resistor that acts as a heater and a 10 k$\Omega$ thermistor that forms part of a voltage-divider circuit (Fig.~\ref{fig:bug}b).  An analog signal from a National Instruments NI-USB-6212 data acquisition device (DAQ) running under NI LabVIEW 2016 is sent to the heater, and the temperature response is read by the same DAQ device.  All signals are sampled at an effective rate of 100~Hz (each point is the average of 4000 points sampled at 400~kHz).  The power signal is measured using a separate NI-USB-6212 DAQ, to minimize crosstalk between the two signals.  Since our analysis is based on linear systems, we nonlinearly transform the output and input voltages into power and temperature signals, respectively.  For power, $P=V^2/R_\text{heater}$; for temperature, we use the standard thermistor calibration curve.  See Ref.~\onlinecite{bechhoefer07} for more details on the experimental setup and thermistor-signal linearization.

In the experimental device, the main element, epoxy, has $\alpha \approx 0.1$ mm$^2$/s and the heater-thermistor distance $\ell \approx 1$ mm.  This leads to a characteristic frequency $f_0 \approx 0.1 / 1^2 / (2\pi) \approx 0.03$ Hz, or a characteristic time $\approx 30$ s.  In order to measure the frequency response, we should explore frequencies well above and below this value.  We chose a logarithmically spaced range of 34 frequencies between $10^{-3}$ and 2 Hz.  

\begin{figure}
\includegraphics[width=3.4in]{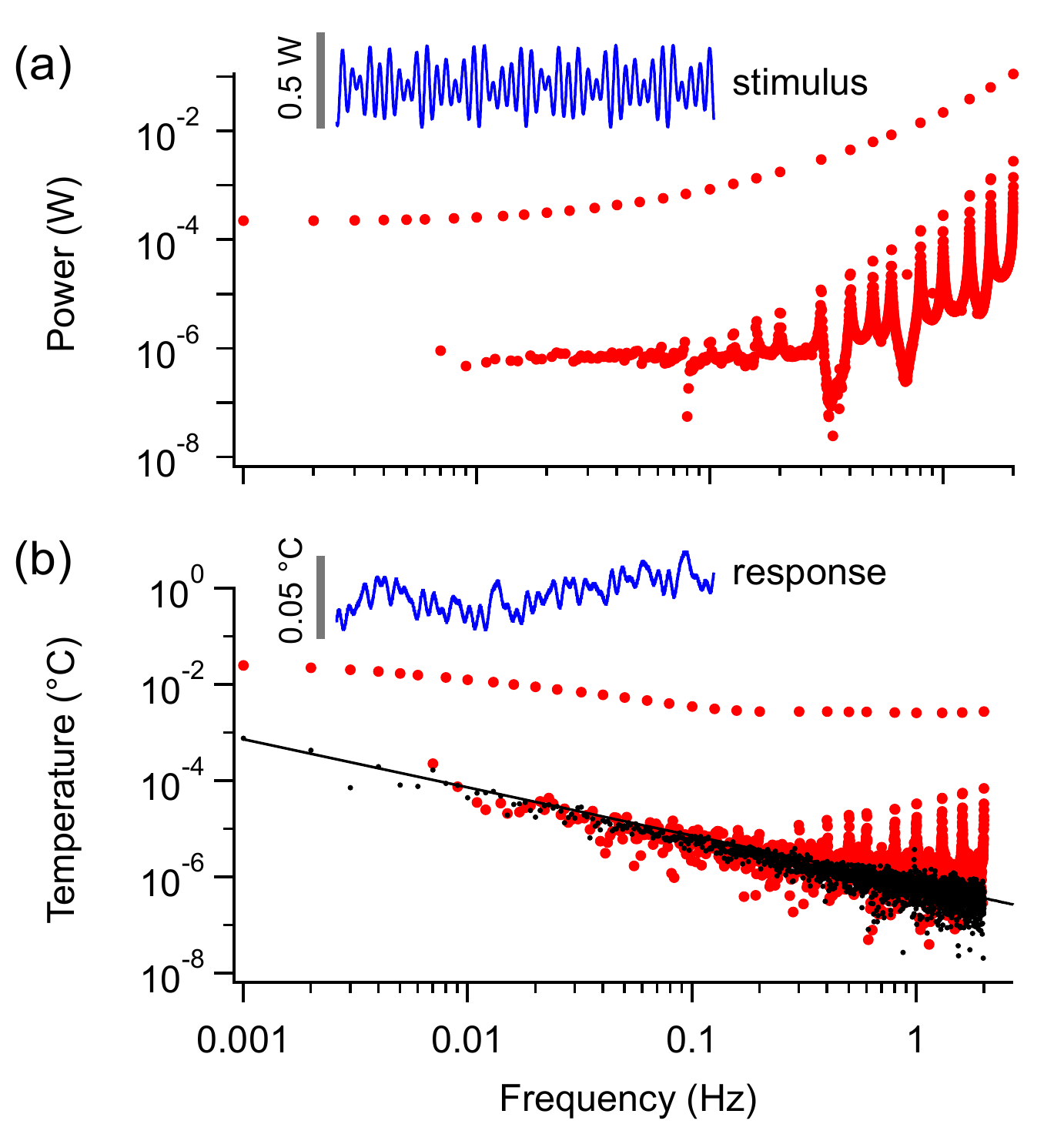}
\caption{Multisine measurement of linear response.  (a)  Amplitudes of the different frequency components of the power signal (stimulus).  Inset (blue) shows 20~s extract of the time-domain excitation signal.  (b)  Amplitudes of the different frequency components of the temperature signal (response).  The overlaid black markers show a measurement of the thermistor noise (amplitude SNR $\gtrsim 100$, or 40 dB at all but the lowest frequencies, where SNR $\approx 10$).  The line is a least-squares fit to a power law with exponent fixed at $-1$.  Inset (blue) shows 20~s of the time-domain response signal corresponding to the inset in (a).}
\label{fig:Multisine}
\end{figure}

With  periods as long as 1000 s, the most straightforward way to measure the response---frequency by frequency---would be very slow.  We therefore implemented the more-efficient multisine method pioneered by Pintelon and Schoukens,\cite{pintelon12} which has only rarely been used in physics applications.  (Reference~\onlinecite{perez15} implements some aspects.)  Taking advantage of linearity, we design a periodic power signal that is a sum of sine waves.  The periodic input signal is,
\begin{equation}
	u(t) = \sum_{n=1}^N A_n \cos (n \omega_0 t + \phi_n) \,,
\label{eq:inputSignal}
\end{equation}
where $\omega_0 = 2\pi f_0$ is the lowest frequency probed, and where we choose the amplitudes $A_n$ and phases $\phi_n$ as discussed below.  Some subtleties to consider:
\begin{itemize}
\item \textit{Frequencies} ($n\omega_0$) are integer multiples of a fundamental.  We record the steady-state response for an integer number of periods of the lowest frequency, to avoid any complications of frequency leakage caused by jumps between the first and last points of the power-signal waveform.  The chosen frequencies are approximately evenly spaced on a logarithmic plot.   Unwanted frequencies have $A_n=0$.

\item  \textit{Amplitudes} ($A_n$) are chosen to be approximately the inverse of the expected response.  (Quick preliminary trials give an adequate estimate of the amplitude response.)  The goal is that the output waveforms should have roughly constant amplitude and, hence, roughly constant signal-to-noise ratios, independent of frequency.  A naive ``white" input signal would lead to temperature fluctuations with negligible high-frequency content.  See Fig.~\ref{fig:Multisine}.

\item \textit{Phases} ($\phi_n$) are chosen randomly.  We tried several sets of random numbers in order to find one that minimized the ``crest factor" of the signal---ratio of peak amplitude to RMS power.  A typical crest factor $\mathcal{C} \approx 1.8$.  For comparison, a sine has $\mathcal{C} = \sqrt{2} \approx 1.4$ and a square wave $\mathcal{C}=1$.   Signals with small crest factors are more efficient at injecting power into the system while minimizing any potential nonlinearities.
\end{itemize}

To measure the response, we then simply record and Fourier analyze the steady-state temperature signal.  The complex ratio of frequency components gives the relative magnitude and phase lag at each frequency.  Examining frequencies in the response that are not present in the stimulus indicate the noise.  They can also indicate system nonlinearities, such as signal linearization.

Some such issues can be seen in the spectral leakage present in Fig.~\ref{fig:Multisine}.  This leakage arose from inaccurate timing arising from hardware limitations of the NI-DAQ-USB-9212, which cannot re-trigger finite acquisitions of data.  While more-sophisticated hardware such as Field-Programmable Gate Arrays (FPGAs) could eliminate these timing issues, we chose to continue our measurements with our DAQ,  because it is typical of the type of instrumentation available to undergraduate teaching laboratories.  Furthermore, the spectral leakage is in the measured \textit{power signal}:  Since the power-temperature response is linear, we care only about the ratio of output to input amplitudes at a particular frequency.  The fact that small amounts of power are injected at other frequencies is irrelevant.  Thus, while the leakage seen here may complicate the measurement of the noise, it does not affect the estimate of the dynamical-response signal.

To isolate noise from both signal and frequency leakage, we measure the thermistor signal while applying only a constant power level, equal to the average power in our measurements.  The result is shown using black markers in Fig.~\ref{fig:Multisine}b, which demonstrates that even at the lowest frequencies measured, the injected signal-to-noise amplitude ratio exceeds 10.  The solid line (black) is a fit to a power law with exponent equal to $-1$, showing that the noise is consistent with the $1/f$ noise that is typical of a resistor with current flowing through it.\cite{weissman88}  At higher frequencies, we would expect a crossover to Johnson thermal (white) noise.  Measuring lower frequencies would be problematic, as the rising $1/f$ noise would start to dominate.

Prior to carrying out the frequency-domain measurements, we remove trends below the minimum frequency in the temperature data by averaging the measurements in each period and fitting a cubic spline through the resulting points.  Subtracting this low-frequency background curve eliminate the effects of a drifting room temperature.  The signal is then segmented into lengths equal to one period of the lowest frequency.  We then Fourier transform each segment and average the complex Fourier amplitudes frequency by frequency.  

\begin{figure}
\includegraphics[width=3.4in]{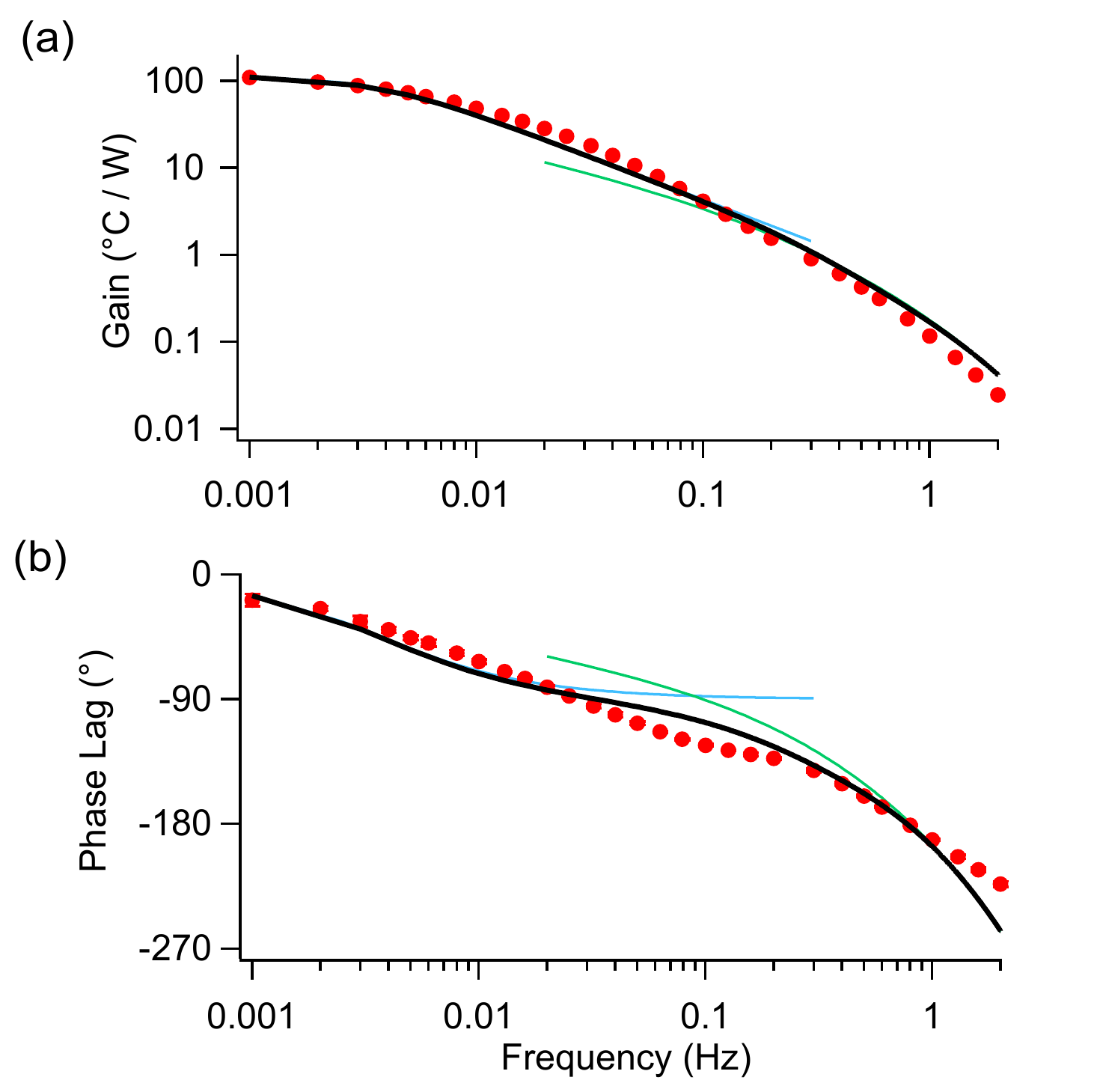}
\caption{Magnitude and phase lag in degrees of the response to a sinusoidally oscillating power input as a function of the frequency of oscillations. The magnitude fits reasonably well while the phase diverges near the crossover between the two limits.  The amplitude model begins to break down for frequencies greater than 1 Hz.}
\label{fig:BodePlots}
\end{figure}

The measured magnitude and phase shift of the temperature response of our system to sinusoidal power oscillations is shown in Fig.~\ref{fig:BodePlots}, along with global fits to the full model, Eq.~\eqref{eq:FullModel} and its low- and high-frequency limits (Eqs.~\ref{eq:lowfreq-Cap} and \ref{eq:HFModel1}).  The most important observation is that the two limits of the model nearly overlap, indicating that the system response is reasonably well described by expressions that are independent of most of the features of the system's geometry.  Recall that we apply an expression for a one-dimensional rod to a blob-shaped, inhomogeneous object.

When formulating the response function, we found it necessary to choose the fit parameters carefully.  Some parameter choices led to degeneracies resulting in large statistical errors in the fit.  We found the following combination worked well:
\begin{subequations}
\begin{alignat}{2}
	K_0 =& \frac{L^2}{\alpha} & &= 5.25 \pm 0.14~\mathrm{s} \,, \\
	K_1 =& \frac{\ell^2}{\alpha} & &= 2.201 \pm 0.016~\mathrm{s} \,, \\
	K_2 =& \frac{A\lambda}{\sqrt{\alpha}} & &= 0.160 \pm 0.002~\mathrm{W\,m/(K\,\sqrt{s})} \,, \\
	K_3 =& \mu & &= 0.0234 \pm 0.0003~\mathrm{s}^{-1} \,.
\end{alignat}
\end{subequations}
All four parameters have straightforward physical interpretations:  The quantity $\sqrt{K_1 / K_0} \equiv \ell / L = 0.65 \pm 0.02$ corresponds to the separation between heater and probe relative to the overall object size.  The inverse of $K_3 \approx 42.7$ s is the time constant for losses to the environment.  The product $\sqrt{K_0} \, K_2 = LA\rho c_p \equiv C = 0.367 \pm 0.014$ J/${}^\circ$C, the heat capacity of the object.  Finally, $K_0$ is the time scale for heat to diffuse across the whole object, while $K_1$ is the time scale for heat to diffuse to the temperature probe.  

The last three quantities combine an intrinsic property (heat capacity $c_p$, thermal diffusivity $\alpha$) with the effective size of the object.  The irregularity of the geometry means that one can make only an approximate estimate of the intrinsic quantities, as the effective lengths are hard to estimate precisely.

We also notice a large deviation from expected phase response at higher frequencies.  This deviation shows the limits to our model:  beyond a maximum frequency, as lengths $\ell_0 \sim \omega^{-1/2}$ decrease, the geometry plays an increasingly important role in the dynamics of the system response.  Surprisingly, the important model deviations affect mostly the phase.  Even when the high-frequency data is excluded from the global fitting, the magnitude of the response naturally fits the model.  

We thus have three experimental regimes:  a lumped-element regime where the system responds as a whole, a one-dimensional regime where our model does a ``good enough'' job, and a high-frequency regime where the response likely needs a full three-dimensional model.

\begin{figure}
\includegraphics[width=3.4in]{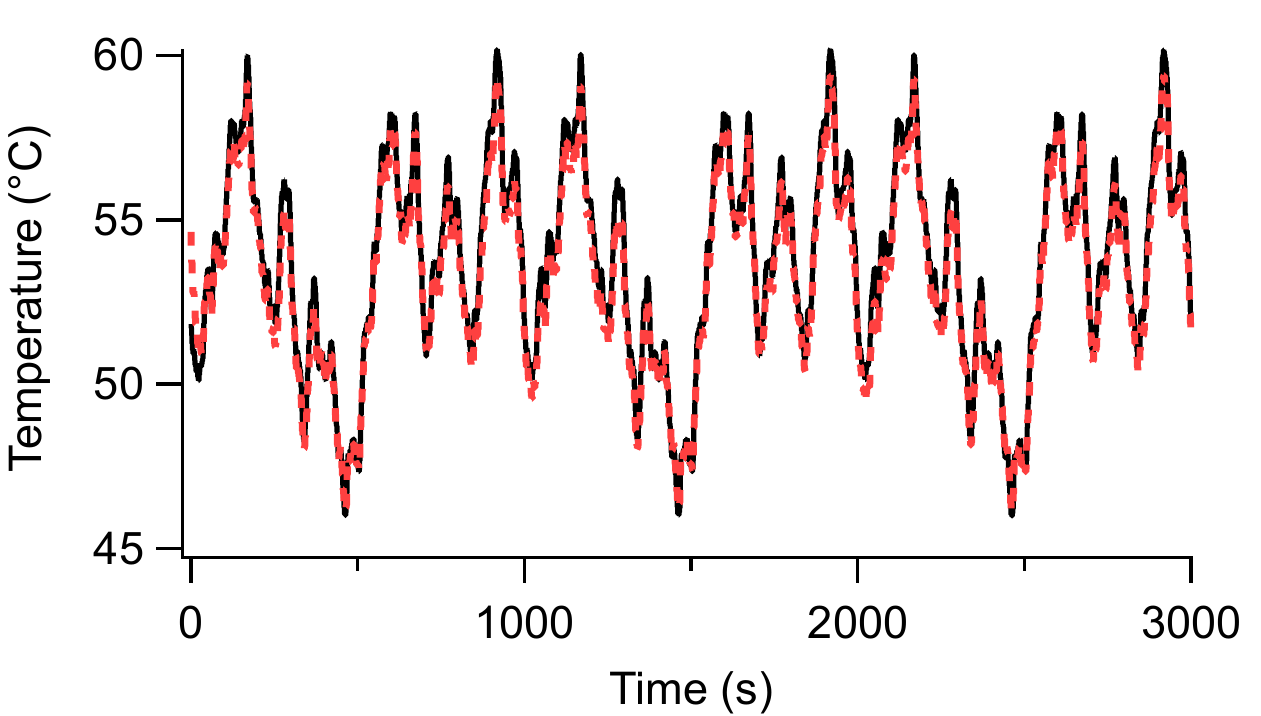}
\caption{Experimental response (solid black line) to a band-limited white signal, a multisine with equal amplitudes at all harmonics and random phases.  Simulation based on model and fit parameters (dashed red line).  Several periods of the pseudo-random signal are shown.}
\label{fig:Sim}
\end{figure}

To further test the performance of our linear-response model, we measured the time-domain response to band-limited ``white noise'' (a periodic signal formed by summing equal-amplitude harmonics with random phases).  The results are shown in Fig.~\ref{fig:Sim}. We compare to numerically generated predictions based on our model and find that the model captures many of the quantitative features;  issues consist mostly of amplitude errors, perhaps indicating the failure of the amplitude at higher frequencies.  Note that both the simulation and recorded data are steady-state response to the repeated input.  It was important to limit the frequencies in the noise input to frequencies over which the model is accurate ($10^{-3}$ to $1$~Hz).  We also note that in order to match the DC component of the temperature, we had to adjust the ambient temperature to $\approx 6~^\circ$C greater than room temperature in our simulations.  This is reasonable, as we had shielded the system from the laboratory environment by placing an insulating Tygon tube around it, capped at the top.  Indeed, measurements of the air temperature inside the shield, which are only a proxy for the ambient temperature of the device, were $\approx 4~^\circ$C higher than the surrounding laboratory temperature.

\section{Conclusion}

We have developed and described a simple model of heat propagation that interpolates between the kind of lumped-element approximations that underlie the circuit theory of elementary physics courses with the simple-geometry partial-differential equations of intermediate mathematical methods courses for physicists.  We found a surprising independence of geometry in the results.  For different reasons, the low- and high-frequency limits do not depend on the detailed shape of the system.  Moreover,  under many circumstances, the domain of validity of both limits nearly overlaps. 

We confirmed these conclusions in a simple experimental system using equipment available in most undergraduate labs, showing that the full model is valid over three decades in frequency for the amplitude and two for the phase.  Time responses show qualitative agreement, with some amplitude and phase errors that lead to systematic shifts.  At higher frequencies in particular, our one-dimensional model breaks down.  A three-dimensional model would likely apply in this region. 

Why is the response so independent of geometrical features?  One reason  is that thermal systems are highly dissipative.  By contrast, in low-loss systems such as light in a cavity or electrons in a box, the behavior is determined by normal modes that intimately reflect the geometry.  More generally, we would also expect that systems with nonlinear, complex dynamics to be poorly described by such simple models.

Beyond the pedagogical value in connecting concepts from elementary- and intermediate-level physics courses, we can gain some insight into the types of ``spherical cow" approximations that physicists are both famous and notorious for using.  Indeed, there are actually two types of ``spherical cows."  The first identifies a simple model that illustrates a physical idea, without too much regard for quantitative agreement with experimental systems.  For these kinds of spherical cows, the exact geometry of an object would be one (of many) details that are neglected in order to illustrate a more basic point.  But ``spherical cow" can have a different sense, where the idea is to  ``engineer" the independence of geometry through careful experimental design.  In this paper, we have given one example of this latter type, where the dissipation associated with diffusion smooths out all but the grossest details of geometry.  Another example is van der Pauw's method for determining the electrical resistivity or Hall coefficient of a thin, irregular object, based on various combinations of current-voltage measurements at four probe points on the border.\cite{vanderpauw58}

In short, the first type of spherical cow abstracts in order to simplify; the second simplifies in order to abstract.  Perhaps we can think of them as the theorist's and experimentalist's versions of spherical cows.

\vspace{2em}

\begin{acknowledgments}

This work was supported by a VP-USRA award from Simon Fraser University to MF, a MITACS GlobalLink fellowship to SG, and an NSERC Discovery Grant to JB.  We thank Jeff Rudd for preparing the sample system.  We thank Mike Hayden, David Broun, and, especially, Steve Dodge for helpful suggestions.
\end{acknowledgments}

\vspace{2em}

\appendix
\section{Response function of a thermal system with distributed losses}
\label{appendix:responseDerivation}

Define the forward Fourier transform as 
\begin{equation}
	\tilde{f}(\omega) = \int_{-\infty}^\infty {\rm d}t \, {\rm e}^{\text{i}\omega t} f(t)  \,.
\label{eq:fourier}
\end{equation}
Then,  Fourier transforming Eq.~\eqref{eq:diffusion} in time leads to
\begin{equation}
	\frac{\partial^2 \tilde{T}}{\partial x^2}-\left(\frac{-\text{i}\omega + \mu}{\alpha} \right) \tilde{T}(\omega,x) = 0,
\end{equation}
with solution
\begin{equation}
	\tilde{T}(\omega,x) = C_1 \, {\rm e}^{ax} + C_2 \, {\rm e}^{-ax} \,,
\label{eq:Temp1}
\end{equation}
where $a = a(\omega) \equiv \sqrt{(\mu-\text{i}\omega)/\alpha}$.  The constants $C_1$ and $C_2$ are set by the boundary conditions in Eq.~\eqref{eq:BCs}.  The condition at $x=L$ implies
\begin{equation}
	\left.\frac{\partial \tilde{T}}{\partial x}\right|_{x=L} = 0 \quad \implies \quad
	C_2=C_1 {\rm e}^{2aL} \,.
\label{eq:bc1}
\end{equation}
The $x=0$ condition, $-\lambda \partial_x \tilde{T} = \tilde{P}/A$, then leads to
\begin{equation}
	\tilde{P}(\omega) = A\lambda a \, C_1 \left( {\rm e}^{2aL}-1 \right)\,.
\label{eq:bc2}
\end{equation}
Combining Eqs.~\eqref{eq:Temp1}, \eqref{eq:bc1}, and \eqref{eq:bc2} gives the linear-response function, $\tilde{\chi}_\ell(\omega) = \tilde{T}(\omega,\ell) / \tilde{P}(\omega)$, of Eq.~\eqref{eq:FullModel}.

\section{Response functions in engineering and physics}
\label{appendix:engVphys}

In the engineering literature, linear response is usually computed in the Laplace, rather than Fourier domain.  To rephrase our result in that language, we substitute $s=-\text{i}\omega$, so that $a(\omega) \to \sqrt{(\mu+s)/\alpha}$ in Eq.~\eqref{eq:FullModel}, and define the transfer function $G(s)$.
 
Beyond the distinction between transfer and response functions, engineers often take a more phenomenological approach to measuring the transfer function.  Using  ``system identification," \cite{ljung99} they treat systems as black boxes and try to find simple dynamical systems that match inputs and outputs accurately.  The resulting transfer functions $G(s)$ are typically rational polynomials in the Laplace variable $s$ and can perform extremely well for a given, fixed system.  They play a key role in the successful feedback control of a system's dynamics.\cite{bechhoefer05,astrom08}  However, they neither generate physical parameters nor can be used when the system changes.  One has to repeat measurements on the new system.



\begin{thebibliography}{99}

\bibitem{bergman11}  T. L. Bergman and A. S. Lavine, \textit{Fundamentals of Heat and Mass Transfer}, 7th ed. (Wiley, 2011), Sections 5.1 and 5.2.

\bibitem{ott88} H.~W.~Ott, \textit{Noise Reduction Techniques in Electronic Systems}, 2nd. ed. (John Wiley \& Sons, 1988).

\bibitem{ramo94}  S.~Ramo, J.~R.~Whinnery, and T.~V.~Duzer, \textit{Fields and Waves in Communication Electronics}, 3rd ed (Wiley, 1994), Ch.~4.

\bibitem{smith97}  H.~J.~T.~Smith and J.~A.~Blackburn, ``Experimental measurements on a simulated lumped transmission line," Am.~J.~Phys.~\textbf{65}, 716--725 (1997).

\bibitem{zangwill13} A.~Zangwill, \textit{Modern Electrodynamics} (Cambridge Univ. Press, 2013), Ch.~14.

\bibitem{landau84}  L.~D.~Landau, E.~M.~Lifshitz, and L.~P.~Pitaevskii, \textit{Electrodynamics of Continuous Media}, 2nd ed. (Pergamon Press, 1984),  Sects. 61--62.

\bibitem{mandelis00} A.~Mandelis, ``Diffusion waves and their uses," Phys. Today \textbf{53}, 29--34 (2000).

\bibitem{ogi16} H.~Ogi, T.~Ishihara, H.~Ishida, A.~Nagkubo, N.~Nakamura, and M.~Hirao, ``Thermal mode spectroscopy for thermal diffusivity of millimeter-size solids," Phys. Rev. Lett. \textbf{117}, 195901 (2016).

\bibitem{zhang17}  J.-C.~Zhang, E.~M.~Levenson-Falk, B.~J.~Ramshaw, D.~A.~Bonn, R.~Liang, W.~N.~Hardy, and S.~A.~Hartnoll, and A.~Kapitulnik, ``Anomalous thermal diffusivity in underdoped YBa$_2$Cu$_3$O$_{6+x}$," Proc. Natl. Acad. Sci. (USA) \textbf{114}, 5378--5383 (2017).

\bibitem{bodas98} A.~Bodas, V.~Gand\'ia, and E.~L\'opez-Baeza, ``An undergraduate experiment on the propagation of thermal waves," Am.~J.~Phys. \textbf{66}, 528--533 (1998).

\bibitem{sullivan08} M.~C.~Sullivan, B.~G.~Thompson, and A.~P.~Williamson, ``An experiment on the dynamics of thermal diffusion," Am.~J.~Phys. \textbf{76}, 637--642 (2008).

\bibitem{anwar14} M.~S.~Anwar, J.~Alam, M.~Wasif, R.~Ullah, S.~Shamim, and W.~Zia, ``Fourier analysis of thermal diffusive waves," Am.~J.~Phys. \textbf{82}, 928--933 (2014).

\bibitem{brody17} J.~Brody and M.~Brown, ``Transient heat conduction in a heat fin," Am.~J.~Phys.~\textbf{85}, 582--586 (2017).

\bibitem{angstrom1861} A.~J.~$\AA$ngstr\"om, ``Neue methode, das {W}\"armeleitungsver-m\"ogen der {K}\"orper zu bestimmen," Annalen der Physik und Chemie \textbf{114}, 513--530 (1861).

\bibitem{carslaw59} H.~S.~Carslaw and J.~C.~Jaeger, \textit{Conduction of Heat in Solids}, 2nd. ed. (Oxford Univ.~Press, 1959).

\bibitem{bechhoefer07}  J. Bechhoefer, Y. Deng, J. Zylberberg, C. Lei, and Z.-G. Ye, ``Temperature dependence of the capacitance of a ferroelectric material," Am. J. Phys 75, 1046--1053 (2007).

\bibitem{pintelon12}  R. Pintelon and J. Schoukens, \textit{System Identification: A Frequency Domain Approach}, 2nd ed. (Wiley IEEE Press, 2012).

\bibitem{perez15} R. P\'erez-Aparicio, C. Crauste-Thibierge, D. Cottinet, M. Tanase, P. Metz, L. Bellon, A. Neart, and S. Ciliberto, ``Simultaneous and accurate measurement of the dielectric constant at many frequencies spanning a wide range," Rev. Sci. Instrum. \textbf{86}, 044702 (2015).

\bibitem{weissman88} M.~B.~Weissman, ``1/f noise and other slow, nonexponential kinetics in condensed matter," Rev.~Mod.~Phys. \textbf{60}, 537--571 (1988).

\bibitem{vanderpauw58} L.~J.~van der Pauw, ``A method of measuring specific resistivity and Hall effect of discs of arbitrary shape," Philips Res. Rep. \textbf{13}, 1--9 (1958). 

\bibitem{ljung99}  L. Ljung, \textit{System Identification: Theory for the User}, 2nd ed.  (Prentice Hall, Upper Saddle River, NJ, 1999).

\bibitem{bechhoefer05}  J. Bechhoefer, ``Feedback for physicists: A tutorial essay on control," Rev. Mod. Phys. \textbf{77}, 783--836 (2005).

\bibitem{astrom08}  K. J. $\AA$str\"om and R. M. Murray, \textit{Feedback Systems: An Introduction for Scientists and Engineers} (Princeton Univ. Press, 2008).

\end{thebibliography}
\end{document}